\documentclass[11pt]{article}
\usepackage{axodraw}
\newcommand{\bmu}{\bar\mu}

\newcommand{\la}[1]{\label{#1}}
\newcommand{\be}{\begin{equation}}
\newcommand{\ee}{\end{equation}}
\newcommand{\ba}{\begin{eqnarray}}
\newcommand{\ea}{\end{eqnarray}}
\newcommand{\rmi}[1]{{\mbox{\scriptsize #1}}}
\newcommand{\fig}{Fig.~}
\newcommand{\figs}{Figs.~}
\newcommand{\eq}{Eq.~}
\newcommand{\se}{Sec.~}
\newcommand{\eqs}{Eqs.~}
\newcommand{\nr}[1]{(\ref{#1})}
\newcommand{\tr}{{\rm Tr\,}}
\newcommand{\nn}{\nonumber \\}
\newcommand{\fr}[2]{{\frac{#1}{#2}\,}}
\newcommand{\msbar}{{\overline{\mbox{\rm MS}}}}

\renewcommand{\vec}[1]{{\bf #1}}
\def\lsi{\raise0.3ex\hbox{$<$\kern-0.75em\raise-1.1ex\hbox{$\sim$}}}
\def\gsi{\raise0.3ex\hbox{$>$\kern-0.75em\raise-1.1ex\hbox{$\sim$}}}
\newcommand{\lsim}{\mathop{\lsi}}

\newcommand{\Tint}[1]{{\hbox{$\sum$}\!\!\!\!\!\!\int}_{\!\!\!\!#1}}

\makeatletter \@addtoreset{equation}{section} \makeatother
\renewcommand{\theequation}{\arabic{section}.\arabic{equation}}
\makeatletter
\renewcommand\section{\@startsection {section}{1}{\z@}%
                                   {-5.5ex \@plus -1ex \@minus -.2ex}
                                   {2.3ex \@plus.2ex}%
                                   {\normalfont\large\bfseries}}
\renewcommand\subsection{\@startsection{subsection}{2}{\z@}%
                                     {-3.25ex\@plus -1ex \@minus -.2ex}%
                                     {1.5ex \@plus .2ex}%
                                     {\normalfont\normalsize\bfseries}}
\renewcommand\thesection {\@arabic\c@section}
\renewcommand\thesubsection   {\thesection.\@arabic\c@subsection}
\renewcommand{\@seccntformat}[1]{%
\csname the#1\endcsname.\hspace{1.0em}}
\makeatother

\begin{document}

\begin{titlepage}
\begin{flushright}
BI-TP 2003/31\\
HIP-2003-58/TH\\
hep-ph/0311268
\end{flushright}
\begin{centering}
\vfill
 
{\Large{\bf Mesonic correlation lengths in high-temperature QCD}}

\vspace{0.8cm}

M. Laine$^{\rm a,}$\footnote{laine@physik.uni-bielefeld.de},
M. Veps\"al\"ainen$^{\rm b,}$\footnote{mtvepsal@pcu.helsinki.fi}

\vspace{0.8cm}

{\em $^{\rm a}$%
Faculty of Physics, University of Bielefeld, 
D-33501 Bielefeld, Germany\\}

\vspace{0.3cm}

{\em $^{\rm b}$%
Theoretical Physics Division, 
Department of Physical Sciences, and Helsinki Institute of Physics,
P.O.Box 64, FIN-00014 University of Helsinki, Finland\\}


\vspace*{0.8cm}
 
\end{centering}
 
\noindent
We consider spatial correlation lengths $\xi$ for various QCD light 
quark bilinears, at temperatures above a few hundred MeV. Some of the
correlation lengths (such as that related to baryon density) coincide
with what has been measured earlier on from glueball-like states;
others do not couple to glueballs, and have a well-known perturbative
leading-order expression as well as a computable next-to-leading-order
correction.  We determine the latter following analogies with
the NRQCD effective theory, used for the study of heavy quarkonia at
zero temperature: we find (for the quenched case) $\xi^{-1} = 2 \pi T
+ 0.1408 g^2 T$, and compare with lattice results.  One
manifestation of U$_\rmi{A}$(1) symmetry non-restoration is also pointed out.
\vfill
\noindent
 

\vspace*{1cm}
 
\noindent
February 2003

\vfill

\end{titlepage}

%
\section{Introduction}
\la{se:introduction}

Given the study of quark--gluon plasma at RHIC and LHC,
there is a continuing need for understanding what QCD actually 
predicts for the various physical observables of interest, in case
the system does reach thermodynamical equilibrium. 
This is a highly non-trivial task, due to the 
strong interactions present in QCD even at temperatures above
a few hundred MeV. Some of the observables can in principle
be determined from first principles using lattice methods, 
but at present systematic errors, related particularly to light 
dynamical quarks, might be substantial. Therefore, 
there is a demand for alternative tools, 
whose results could be compared
with those from lattice simulations.  

Because of asymptotic freedom, a natural candidate for such a tool
is perturbation theory. Alas, it suffers from serious infrared (IR) 
problems at high temperatures~\cite{linde,gpy}. These problems are 
of a rather specific kind, however: they concern directly only 
static, long-range gluons. This has lead to the general 
framework of dimensionally reduced effective field theories~\cite{dr}.
The idea is that there are still a number of computations which 
can be carried out reliably in perturbation theory, while the parts
which cannot, can be treated with lattice simulations using an 
effective theory which  is
much simpler than the original one, all the quarks
having been integrated out. 

During the last few years, the program of dimensional reduction 
has been applied to a large variety of physical quantities: 
in particular, to various gluonic correlation 
lengths~\cite{drold}--\cite{ckp} 
as well as the equation of state~\cite{bn}--\cite{tr03}. 
Comparisons with 
4d lattice results in pure SU(2) and SU(3) Yang-Mills theories
are summarised in~\cite{oweal,mu,sewm03,lprev,chris}, and they suggest
that dimensional reduction is applicable as soon as the temperature 
is somewhat above $T_c$, where $T_c$ is the critical temperature for the
deconfinement phase transition.

An even simpler pattern can perhaps be argued for~\cite{a0cond,sewm03,chris}:
it appears, inspecting the empirical results from the studies
mentioned, that a truncated coupling constant series may already give
a reasonable part of the full answer, if carried out to a sufficient 
depth such that all dynamical scales in the system have made their entrances. 
Therefore, either non-perturbative studies with 
the dimensionally reduced theory, or the further
reduction suggested by this prescription, 
may provide a welcome alternative to direct 4d lattice 
simulations of QCD, particularly for observables where the 
systematic errors are not well under control. 
It should be mentioned that
other proposals for analytic tools exist, as well
(for recent reviews, see~\cite{rvs}).

An important class of observables which have attracted
somewhat less attention in these settings, are those
built out of quark fields, rather than gluonic fields. In general, 
they are expected to be less IR sensitive and thus more perturbative
than purely gluonic observables, so that analytic tools should be better 
applicable. At the same time, there is more urgency for analytic
tools, given the severe difficulties in treating light 
dynamical fermions on the lattice. On top of these formal 
motivations, ``mesonic'' and ``baryonic'' observables may be more directly
related to experimental signatures than gluonic ones. During
the last few years, significant progress has been made
on one class of observables of this type, that is various
quark number susceptibilities 
(see, e.g.,~\cite{oldsus}--\cite{av}). 
The purpose of this paper is to address another class, closely related
but more directly sensitive to IR physics: the long-distance
spatial correlation lengths related to mesonic operators
(i.e., gauge-invariant quark bilinears)
constructed out of the light quark flavours~\cite{dtk}. 

There is a large variety of independent mesonic observables 
of the type mentioned.
Assuming, as we will in this paper, QCD to consist of $N_f$ flavours
of degenerate quarks, they can be divided, first of all, into singlets
and non-singlets under the flavour group.
Each class can furthermore be divided into scalar, pseudoscalar, 
vector, and axial vector objects. Physically perhaps the most interesting
quark bilinears are those representing baryon number density 
and electric charge density, but since all the other ones can 
in principle be measured on the lattice equally well
and this is the most immediate reference point for our 
results, we will try to be as general as possible.   

Previous analytic work on similar topics was 
initiated more than a decade ago~\cite{hz}--\cite{ish}, 
following the first numerical studies~\cite{dtk}. 
(There was even earlier work for small
temperatures, $0 < T \ll T_c$~\cite{gl,ei}.) 
Many of these studies had mostly
qualitative motivations, however: corrections of order
unity with respect to the ones computed 
were (knowingly) left out. They were also partly motivated by 
some qualitative features in the early lattice results
(particularly that the vector channel, or $\rho$, appeared to 
be significantly ``heavier'' than the pseudoscalar channel, or $\pi$, 
even at $T > T_c$),
which have turned out to be 
finite lattice spacing artifacts. 
Significant further progress was made in~\cite{hl}, and we show here, 
following that line,
that the corrections neglected are numerically important, and even modify 
the parametric form of the final result, by removing any logarithms. 
Note that a somewhat related
problem has been addressed in the context of the 
electroweak theory at finite temperatures~\cite{al}, 
although for operators which are bilinear in the Higgs field 
rather than in fermionic fields.

The plan of this paper is the following. 
The basic notation used throughout is fixed in~\se\ref{se:basic}.
Some well-known properties of mesonic correlators  
at very high temperatures are reviewed in~\se\ref{se:limits}. 
In~\se\ref{se:nrqcd} we discuss how
the mesonic correlators can be incorporated into the general 
setting of describing high-temperature QCD by a dimensionally 
reduced effective field theory~\cite{hl}. The quark part of the 
Lagrangian entering will be called three-dimensional 
(3d) ``Non-Relativistic QCD'', NRQCD$_3$. The basic matching 
computation between the original QCD and NRQCD$_3$
is carried out in~\se\ref{se:match}, and 
the first non-trivial correction
to the correlation length of flavour non-singlet mesons
is then determined from NRQCD$_3$ in~\se\ref{se:potential}. 
In~\se\ref{se:result} we 
compare with lattice results. Flavour 
singlet observables, for which the pattern is
qualitatively different from the non-singlets,  
since they may mix with purely gluonic operators, 
are treated in~\se\ref{se:singlets}, 
and we conclude in~\se\ref{se:concl}.

%
\section{Notation}
\la{se:basic}

The theory we consider in this paper is QCD at a finite temperature $T$, 
with the Euclidean quark Lagrangian 
\be
 \mathcal{L}_E^\psi = 
 \bar \psi (\gamma_\mu D_\mu + M )\psi
 \;, 
\ee
where $D_\mu = \partial_\mu - i g A_\mu$, 
$A_\mu = A_\mu^c T^c$, ${T}^c$
are Hermitean generators of SU($N_c$) normalised such that 
$\tr T^c T^d = \fr12 \delta^{cd}$, 
and $M$ is the mass matrix. For simplicity, we take $M$ to be 
diagonal and degenerate, $M = \mathop{\mbox{diag}}(m,...,m)$, 
and in most of what follows, $m=0$. The number of flavours 
appearing in $\psi$ is denoted by $N_f$. 

Given $\bar\psi, \psi$ we may define, as usual, various
bilinears. The 
$N_f \times N_f$ flavour basis is generated by 
\be
 F^a \equiv \{ F^s, F^{n} \}, \quad
 F^s \equiv 1_{N_f\times N_f}, \quad {n} = 1,...,N_f^2-1
 \;,
\ee
where $F^s = {1_{N_f\times N_f}}$
is an $N_f\times N_f$ identity matrix (``singlet''),  and
the traceless $F^{n}$ (``non-singlet'') are assumed normalised such that 
\be
 \tr [F^{m} F^{n}] = \fr12 \delta^{mn} 
 \;. 
 \la{Fnorm}
\ee
We may then consider scalar, 
pseudoscalar, vector, and axial vector objects, 
\ba
 S^{a} & \equiv & \bar\psi F^{a} \psi 
 \;, \la{Sa} \\
 P^{a} & \equiv & \bar\psi \gamma_5 F^{a} \psi 
 \;, \la{Pa} \\
 V^{a}_\mu & \equiv & \bar\psi \gamma_\mu F^{a} \psi
 \;, \la{Va} \\
 A^{a}_\mu & \equiv & \bar\psi \gamma_\mu \gamma_5 F^{a} \psi
 \;. \la{Aa}
\ea

The operators just defined may be interpreted either directly 
as physical condensates or currents, or as interpolating operators for 
certain particle states. In three-flavour QCD ($N_f = 3$) with 
a degenerate mass matrix, for instance, some possible assignments
are shown in~Table~\ref{ta:rev}.

\begin{table}[t]

\begin{center}
\begin{tabular}{llll}  
\hline\hline
Operator        & Particle           
                            & Behaviour in free theory
          & Physical interpretation         \\ 
\hline\hline
~~$S^s$       & $f_0/\sigma$  
                            & $q^2 B_\rmi{3d}$($2 p_0$) & \\
~~$S^n$       & $a_0/\delta$  
                            & $q^2 B_\rmi{3d}$($2 p_0$) & \\
~~$P^s$       & $\eta'$       
                            & $q^2 B_\rmi{3d}$($2 p_0$) & \\
~~$P^n$       & $\pi,\eta$    
                            & $q^2 B_\rmi{3d}$($2 p_0$) & \\
~~$V^s_0$     & $\omega$      
                            & $({q}^2 + 4 p_0^2)$$B_\rmi{3d}$($2 p_0$)
 & baryon density \\
~~$V^s_\perp$ &               
                            & $({q}^2 - 4 p_0^2)$$B_\rmi{3d}$($2 p_0$) & \\
~~$V^n_0$     & $\rho,\phi$   
                            & $({q}^2 + 4 p_0^2)$$B_\rmi{3d}$($2 p_0$)
 & charge density \\
~~$V^n_\perp$ &               
                            & $({q}^2 - 4 p_0^2)$$B_\rmi{3d}$($2 p_0$) & \\
~~$A^s_0$     & $f_1$         
                       & $({q}^2 + 4 p_0^2)$$B_\rmi{3d}$($2 p_0$)  & \\
~~$A^s_\perp$ &               
                            & $({q}^2 - 4 p_0^2)$$B_\rmi{3d}$($2 p_0$) & \\
~~$A^n_0$     & $a_1$       
                     & $({q}^2 + 4 p_0^2)$$B_\rmi{3d}$($2 p_0$) & \\
~~$A^n_\perp$ &              
                            & $({q}^2 - 4 p_0^2)$$B_\rmi{3d}$($2 p_0$) & \\
\hline\hline
\end{tabular}
\end{center}

\caption[a]{Different particle assignments and
the nature of the singularity structure for
the mesonic correlators considered (cf.\ \se\ref{se:limits}). 
The physical interpretation
of $V_0^n$ applies for $N_f = 3$. The notation $(...)_\perp$
refers to polarizations transverse with respect to the spatial direction
in which the correlation is measured; the longitudinal correlators
all vanish in perturbation theory, due to current conservation
in the chiral limit.}
\la{ta:rev}

\end{table}

Given such operators, the correlators to be considered are of the form 
\ba
 C_\vec{q}[O^{a},O^{b}] \equiv
 \int_0^{1/T} {\rm d}\tau
 \int {\rm d}^3 x\, e^{i \vec{q}\cdot \vec{x}} \,
 \langle
 O^{a} (\tau,\vec{x}) O^{b}(0,\vec{0}) 
 \rangle
 \;.
 \la{Cq}
\ea
The corresponding configuration space correlator is
\ba
 C_\vec{x}[O^{a},O^{b}] \equiv
 \int\! \frac{{\rm d}^3 q }{(2\pi)^3}\, e^{-i \vec{q}\cdot \vec{x}} \,
 C_\vec{q}[O^{a},O^{b}] = 
 \int_0^{1/T} \! {\rm d}\tau \, 
 \langle O^a(\tau,\vec{x}) O^b(0,\vec{0}) \rangle
 \;.
 \la{space} \la{Cx}
\ea
The expectation values are taken in a volume which is infinite in 
the spatial directions but finite in the temporal direction, with 
extent $1/T$. Gauge fields are assumed to obey periodic, fermions
anti-periodic boundary conditions around the temporal direction.

The general structures of the correlators so defined are, 
because of rotational invariance, 
\ba
 C_\vec{q}[S^{a},S^{b}] \;, 
 C_\vec{q}[V^{a}_0,V^{b}_0] & = & \delta^{{a}{b}} f(q^2) 
 \;, \\
 C_\vec{q}[V^{a}_i,V^{b}_j] & = & \delta^{{a}{b}} 
 \Bigl[ \Bigl( \delta_{ij} - \frac{q_i q_j}{{q}^2} \Bigr) t(q^2) +
 \frac{q_i q_j}{{q}^2} l(q^2)
 \Bigr]
 \;, 
\ea
and correspondingly with $S\to P$, $V \to A$. Here $q \equiv |\vec{q}|$. 
For conserved currents, $l(q^2) = 0$.

We expect, in general, that in the full 
theory the analytic structures of the functions
$f(q^2),t(q^2),l(q^2)$ around the origin are 
characterised by simple poles. The reason  
is that, as we will recall presently, these functions can 
after analytic continuation be viewed as propagators of 
physical states in a (2+1)-dimensional confining theory. 
The spectrum of the confining theory consists of bound
states, represented by poles. Consequently, 
the Fourier transforms should show exponential falloff
at large distances. 
The purpose of this paper is to determine 
the pole locations or, equivalently, the long-distance 
exponential falloffs of the corresponding configuration
space correlation functions, to order $\mathcal{O}(g^2)$. 
The coefficients of the exponential
falloffs are often referred to as screening masses.

%
\section{Correlators at very high temperatures}
\la{se:limits}

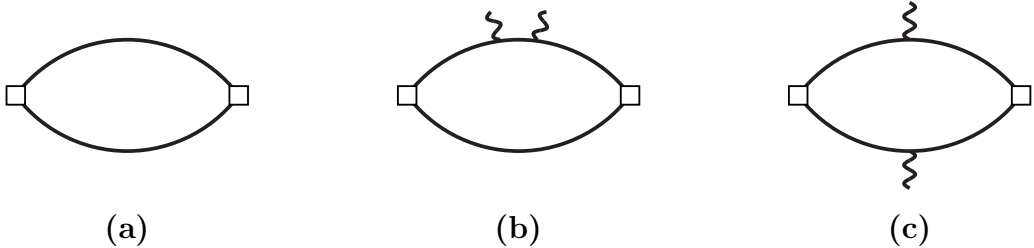
\begin{figure}[t]

\def\TAsc(#1,#2)(#3,#4,#5)%
{\SetWidth{2.0}\CArc(#1,#2)(#3,#4,#5)\SetWidth{1.0}}
\def\Lwidth{3}

\def\TAgl(#1,#2)(#3,#4,#5){\SetWidth{2.0}\PhotonArc(#1,#2)(#3,#4,#5){\Lwidth}%
{6.283 #3 mul 360 div #4 #5 sub #4 #5 sub mul sqrt mul Tdensity mul}%
\SetWidth{1.0}}
\def\TLgl(#1,#2)(#3,#4){\SetWidth{2.0}\Photon(#1,#2)(#3,#4){\Lwidth}
{#1 #3 sub #1 #3 sub mul #2 #4 sub #2 #4 sub mul add sqrt Tdensity mul}%
\SetWidth{1.0}}
\newcommand{\piC}[1]{\;\parbox[c]{120pt}{\begin{picture}(120,60)(0,-20)
\SetWidth{1.0}\SetScale{0.7} #1 \end{picture}}\;}
\def\Treelevel(#1,#2){\piC{#1(60,-15)(75,34,146) #2(60,75)(75,214,326)%
\GBoxc(0,30)(10,10){1} \GBoxc(120,30)(10,10){1}%
\Text(42,-30)[c]{\large\bf (a)} }}
\def\Sameside(#1,#2,#3,#4){\piC{#1(60,-15)(75,34,146) #2(60,75)(75,214,326)%
 #3(50,60)(45,75) #4(70,60)(75,75)%
\GBoxc(0,30)(10,10){1} \GBoxc(120,30)(10,10){1}%
\Text(42,-30)[c]{\large\bf (b)} }}
\def\Opposite(#1,#2,#3,#4){\piC{#1(60,-15)(75,34,146) #2(60,75)(75,214,326)%
 #3(60,60)(60,80) #4(60,0)(60,-20)%
\GBoxc(0,30)(10,10){1} \GBoxc(120,30)(10,10){1}%
\Text(42,-30)[c]{\large\bf (c)} }}

\vspace*{0.5cm}

\begin{eqnarray*}
 & & 
 \Treelevel(\TAsc,\TAsc) \qquad 
 \Sameside(\TAsc,\TAsc,\TLgl,\TLgl) \qquad 
 \Opposite(\TAsc,\TAsc,\TLgl,\TLgl) 
\end{eqnarray*}

\vspace*{0.4cm}

\caption[a]{\it (a): The leading order correlator. (b), (c):
Graphs representing non-trivial gluon condensates in the operator 
product expansion for the mesonic correlation function.}  
\label{fig:lo}

\end{figure}

To start with, it may be useful to recall the behaviour
of the correlation functions we are interested in
at very high temperatures. 
In this limit, asymptotic freedom should guarantee that 
we can use perturbation theory, and the correlators can be computed
with free fermions (\fig\ref{fig:lo}(a)). 
One finds, in dimensional regularisation, 
\be
 C_\vec{q}[O^a,O^b] = 
 \tr [ F^{a} F^{b} ] \, N_c \,
 T \sum_{n=-\infty}^{\infty} 
 \int \frac{{\rm d}^{3-2\epsilon} p}{(2\pi)^{3-2\epsilon}}
 \frac{1}{[p_n^2 + \vec{p}^2][p_n^2 + (\vec{p}+ \vec{q})^2]}
 \tr \Bigl[ 
 (\slash\!\!\! p + \slash\!\!\! q)\, 
 \Gamma^{a} \, \slash\!\!\! p \, \Gamma^{b}
 \Bigr]
 \;,
 \la{free}
\ee
where $\Gamma^{a}$ is the Dirac matrix appearing in 
$O^a$, $\Gamma^{a} = \{1,\gamma_5,\gamma_\mu,\gamma_\mu \gamma_5\}$, 
$\slash\!\!\! p \equiv \gamma_\mu p_\mu$, 
and $p_n$ denotes the fermionic 
Matsubara frequencies, $p_n  \equiv 2 \pi T (n + \fr12 )$.
The Dirac algebra and the integration are trivially carried out: 
the result contains either
constants, which correspond to contact terms $\sim \delta(\vec{x})$
after going into configuration space according to~\eq\nr{space}, 
or the function
\be
 B_\rmi{3d}(2 p_n) \equiv 
 \int \frac{{\rm d}^3 p}{(2\pi)^3}
 \frac{1}{[p_n^2 + \vec{p}^2 ][p_n^2 + (\vec{p}+\vec{q})^2 ]} = 
 \frac{i}{8\pi q} \ln\frac{2 p_n - i q}{2 p_n + i q} \;.
 \la{B3d}
\ee
The qualitative functional forms of the terms including  this
function for the different $O^a$ are shown in~Table~\ref{ta:rev}, 
for $p_n  = p_0$.

The expression in~\eq\nr{free} contains 
a sum over all $p_n  = 2 \pi T(n + \fr12)$. Since
the non-trivial structure appears at the point $2 p_n$
(cf.\ \eq\nr{B3d}), the correlator is dominated at large distances
(small ${q}$)
by the smallest Matsubara frequencies, $\pm p_0 = \pm \pi T$. 
These frequency modes live in three spatial dimensions. If we now imagine
viewing the three-dimensional theory rather as a (2+1)-dimensional one, and 
call the direction in which the correlation is measured, the time $t$, 
then the screening mass corresponds to the energy of a state consisting
of two on-shell massive quarks, each of ``mass'' $p_0$. 
Put in this language, our problem will be 
to compute the first correction to the energy
of the two-particle state. 

We may also note that the nature of the singularity
at $q = \pm 2 i p_0$ in~\eq\nr{B3d} 
is a branch cut. In a (2+1)-dimensional world, this just corresponds
to an energy threshold for producing
two free on-shell particles of ``mass'' $p_0$.  We may expect this structure 
to get qualitatively modified by the interactions, which
place the quarks off-shell and bind them together, as just described:
the threshold  should then convert to a pole.
The computation of the functional form of the singularity is beyond
the scope of the present paper, however: we simply study how
the location of the threshold at $2 p_0$ gets modified. 

Note from Table~\ref{ta:rev}
that for the charge-charge correlators ($V^a_0,A^a_0$), 
there is a prefactor $({q}^2 + 4 p_0^2)$ which vanishes at the point
of the branch singularity in $B_\rmi{3d}(2 p_0)$. 
This means that a contribution of the singularity at 
$q = \pm 2 i p_0$ is suppressed, and 
the correlator tends to decay even faster:
it is easy to see that in configuration space,
it is power-suppressed in $1/p_0 t$.
The issue is even clearer if, for a moment, 
one imagines going to a 
world with one space dimension only: then 
\be
 B_\rmi{1d}(2 p_0) = 
 \int \frac{{\rm d} p}{2\pi}
 \frac{1}{[p_0^2 + p^2 ][p_0^2 + (p+q)^2 ]} = 
 \frac{1}{p_0} \frac{1}{q^2 + 4 p_0^2} 
 \;, 
\ee
and the singularity (a real pole in this case) would be completely
cancelled by the prefactor, leaving only (short-distance)
contact terms\footnote{%
  This remains true even after inclusion of the other $p_n$'s.}. 
We will return to this issue below.

To conclude this Section, we briefly inspect the correlators from 
yet another angle. Let us, for a moment, consider the correlator  
not around the pole at $q = \pm 2 i p_0$, but for a very large
(real) ${q}$. Then the correlator can be computed 
in the operator product expansion: taking a background of constant 
(zero Matsubara mode) gauge fields (\figs\ref{fig:lo}(b), (c))
one obtains, for instance,
\be
 C_\vec{q}[S^a,S^b] \approx 
 \tr [ F^a F^b ] \, T \biggl[ 
  - \frac{N_c}{4\pi} \biggl(
 4 p_0 + i q \ln\frac{2 p_0 - i q}{2 p_0 + i q} 
 \biggr) 
 + \frac{1}{\pi p_0}
 \frac{(q^2)^2}{(q^2 + 4 p_0^2)^2} g^2 \tr A_0^2 + ... 
 \biggr]
 \;.
 \la{ope}
\ee
Extrapolating then back towards the threshold region, 
say $q \sim 2 i p_0 + \mathcal{O}(g^2T)$, 
one sees how the corrections related to the gauge fields 
are becoming increasingly important. In order to systematically
evaluate them we have to turn, however, to other methods. 

%
\section{Basic structure of NRQCD$_3$}
\la{se:nrqcd}

\begin{figure}[t]

\def\TAsc(#1,#2)(#3,#4,#5)%
{\SetWidth{2.0}\CArc(#1,#2)(#3,#4,#5)\SetWidth{1.0}}
\def\Lwidth{3}

\def\TAgl(#1,#2)(#3,#4,#5){\SetWidth{2.0}\PhotonArc(#1,#2)(#3,#4,#5){\Lwidth}%
{6.283 #3 mul 360 div #4 #5 sub #4 #5 sub mul sqrt mul Tdensity mul}%
\SetWidth{1.0}}
\def\TLgl(#1,#2)(#3,#4){\SetWidth{2.0}\Photon(#1,#2)(#3,#4){\Lwidth}
{#1 #3 sub #1 #3 sub mul #2 #4 sub #2 #4 sub mul add sqrt Tdensity mul}%
\SetWidth{1.0}}
\newcommand{\piC}[1]{\;\parbox[c]{120pt}{\begin{picture}(120,60)(0,-20)
\SetWidth{1.0}\SetScale{0.7} #1 \end{picture}}\;}
\def\ConnectedA(#1,#2,#3){\piC{#1(60,-15)(75,34,146) #2(60,75)(75,214,326)%
 #3(60,60)(20,190,350)%
\GBoxc(0,30)(10,10){1} \GBoxc(120,30)(10,10){1}%
\Text(42,-15)[c]{\large\bf (a)} }}
\def\ConnectedB(#1,#2,#3){\piC{#1(60,-15)(75,34,146) #2(60,75)(75,214,326)%
 #3(60,60)(60,0)%
\GBoxc(0,30)(10,10){1} \GBoxc(120,30)(10,10){1}%
\Text(42,-15)[c]{\large\bf (b)} }}
\def\Disconnected(#1,#2,#3,#4){\piC{#1(25,30)(25,183,177) #2(95,30)(25,0,360)%
\SetWidth{2.0}\Photon(48,40)(72,40){-3}{2.5}\SetWidth{1.0} %
\SetWidth{2.0}\Photon(72,20)(48,20){-3}{2.5}\SetWidth{1.0} %
\GBoxc(0,30)(10,10){1} \GBoxc(120,30)(10,10){1}%
\Text(42,-15)[c]{\large\bf (c)} }}

\begin{eqnarray*}
 & & 
 \ConnectedA(\TAsc,\TAsc,\TAgl) \qquad 
 \ConnectedB(\TAsc,\TAsc,\TLgl) \qquad 
 \Disconnected(\TAsc,\TAsc,\TLgl,\TLgl) 
\end{eqnarray*}

\caption[a]{\it Schematic representations of various classes of 
beyond-the-leading order corrections: (a) quark
self-energy correction; (b) quark--antiquark interaction via
gluon exchange; (c) for flavour singlets, the correlation 
can also be mediated by purely gluonic states.}  
\label{fig:nrt}

\end{figure}
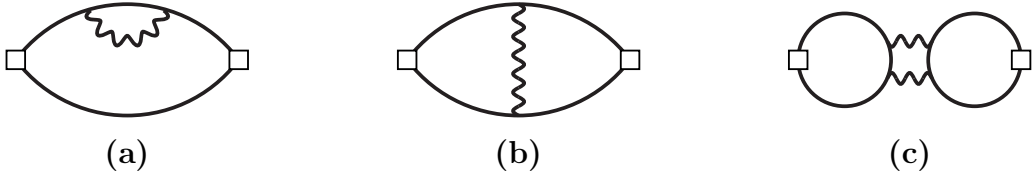

To go beyond the accuracy of \fig\ref{fig:lo}(a), we need 
to compute graphs of the qualitative types in~\fig\ref{fig:nrt}. 
Just the graphs shown are not enough, however: one also needs 
to account for various kinds of iterations of these basic topologies. 
The reason is that near the threshold, quarks can be almost on-shell,  
$|1/\slash\!\!\! p|\sim \mathcal{O}(1/g^2T)$, and thus compensate for
the appearance of $\mathcal{O}(g^2T)$ in the vertices. Therefore, there 
is a large set of graphs which has to be identified and resummed, in 
order to obtain the first correction to the threshold location. 

As in many other contexts, a convenient organising principle for
carrying out such resummations is offered by effective field theory
methods. As already explained, 
in the present case we can view the correlation lengths  
as (2+1)-dimensional bound states of heavy particles of ``mass'' $p_0$, 
much larger than the infrared scales $\sim gT,g^2T$ of gauge field
dynamics at high temperatures~\cite{dr}. The masses of such heavy 
bound states can be determined, like those of quarkonia at $T=0$, through 
an effective theory called the non-relativistic QCD (NRQCD)~\cite{nrqcd}.
Our case does lead to some significant differences with 
respect to this framework; for instance, that the ``masses'', 
$p_0$, are not really masses in the conventional sense, but conserve chiral
invariance and are directly scale invariant physical quantities. 
Despite the differences, we will refer to the general framework  
to be used as NRQCD$_3$. The philosophy of applying
NRQCD techniques to the present problem was introduced by 
Huang and Lissia~\cite{hl}. We start by showing how 
the non-relativistic structure emerges on the tree-level. 

\subsection{Tree-level}
\la{se:tree}

As seen in~\eq\nr{free}, at leading order the correlators
considered get independent contributions from the various 
quark Matsubara modes, with the lowest ones being dominant at 
large distances. Let us therefore
again concentrate on $p_0 \equiv \pi T$.
The quark Lagrangian for this Matsubara mode is
\be
 \mathcal{L}_E^\psi = 
 \bar\psi
 \Bigl[ 
 i \gamma_0 p_0 
 - i g \gamma_0 A_0 + 
 \gamma_k D_k + \gamma_3 D_3
 \Bigr]
 \psi
 \;,
 \la{Lpsi}
\ee 
where $k\equiv 1,2$. 

In \eq\nr{Lpsi}, $\psi$ interacts 
with gluonic zero Matsubara modes only. The
Lagrangian describing their dynamics, obtained after
integrating out all quarks and the non-zero Matsubara
modes of gluons, is the dimensionally reduced Lagrangian, 
of the form~\cite{dr} 
\be
 \mathcal{L}_E^A = 
 \fr12 \tr F_{ij}^2 + \tr [D_i,A_0]^2 + 
 m_\rmi{E}^2\tr A_0^2 +\lambda_\rmi{E}^{(1)} (\tr A_0^2)^2
 +\lambda_\rmi{E}^{(2)} \tr A_0^4 + ... \;.
 \; 
 \la{LA}
\ee
Here $i=1,...,3$, $F_{ij} = (i/g_\rmi{E}) [D_i,D_j]$, 
and $D_i = \partial_i - i g_\rmi{E} A_i$, 
where $A_\mu = A_\mu^c {T}^c$ and the fields 
have now the dimension GeV$^{1/2}$, after a trivial
rescaling with $T^{1/2}$.
The parameters, to the orders that we need them, are 
\be
 m_\rmi{E}^2 = \Bigl(\frac{N_c}{3} + \frac{N_f}{6}\Bigr) g^2T^2
 \;,
 \quad 
 g_\rmi{E}^2 = g^2T
 \;,
 \la{prms} 
\ee
while $\lambda^{(i)}_\rmi{E} = \mathcal{O}(g^4T)$ can be ignored.
The parameters have been determined up to 
next-to-leading order~\cite{hl1,bn,generic}. 
According to \eqs\nr{LA}, \nr{prms}, 
infrared gauge field dynamics is sensitive to two 
scales: the perturbative scale $m_\rmi{E} \sim gT$
associated with the colour-electric component $A_0$, 
and the (2+1)-dimensional 
confinement scale $g_\rmi{E}^2 \sim g^2 T$
associated with the colour-magnetic components $A_i$.

Let us next set up some notation. In the following, we use rotational
invariance to choose the vector $\vec{q}$ in~\eq\nr{Cq} to point in
the $x_3$-direction. According to~\eq\nr{space} we are thus
effectively considering correlators averaged over the
$(x_1,x_2)$-plane, and depending on $x_3$. We will often 
denote the Euclidean $x_3$-coordinate by $t$, 
and the vector $(x_1,x_2)$ by $\vec{x}_\perp$. 
The corresponding Fourier modes 
are $p_3,\vec{p}_\perp$. The indices $k,l$ are 
assumed to have the values $1,2$, and label the transverse directions. 
Given this choice of $\vec{q}$ we can consider, without loss
of generality, the correlator
\be
 C_{t}[O^{a},O^{b}] \equiv
 \int\! \frac{{\rm d} q_3 }{2\pi}\, e^{-i q_3 t} \,
 C_{(0,0,q_3)}[O^{a},O^{b}] = 
 \int_0^{1/T} \! {\rm d}\tau \, \int \!{\rm d}^2 \vec{x}_\perp\,
 \langle O^a(\tau,\vec{x}_\perp,t) O^b(0,\vec{0}_\perp,0) \rangle
 \;,
 \la{Ct}
\ee
rather than $C_\vec{x}[O^a,O^b]$ in \eq\nr{Cx}.

Once this choice has been made, 
the quark Lagrangian in~\eq\nr{Lpsi} 
can be written in a suggestive form. 
Let us make a basis transformation from the 
standard representation 
\be
 \gamma_0 = 
 \left( 
 \begin{tabular}{cc}
 $1_{2\times 2}$ & 0 \\ 
 0 & $-1_{2\times 2}$ \\ 
 \end{tabular}
 \right) \;, \quad
 \gamma_i = 
 \left( 
 \begin{tabular}{cc}
 0 & $-i\sigma_i$ \\ 
 $i\sigma_i$ & 0  \\ 
 \end{tabular}
 \right) \;, \quad
 \gamma_5 = 
 \left( 
 \begin{tabular}{cc}
 0 & $1_{2\times 2}$ \\ 
 $1_{2\times 2}$ & 0 \\ 
 \end{tabular}
 \right) \;, \quad
\ee
where $\sigma_i$ are the Pauli matrices,
to a new basis, $\gamma^\rmi{new}_\mu = 
U \gamma^\rmi{standard}_\mu U^{-1}$, with
\be
 U = \frac{1}{\sqrt{2}}
 \left( 
 \begin{tabular}{cccc}
 1 & 0 & -1 &  0 \\ 
 0 & 1 &  0 &  1 \\
 1 & 0 &  1 &  0 \\
 0 & 1 &  0 & -1 
 \end{tabular} 
 \right) \;.
\ee
In this basis, 
\be
 \gamma_0 \slash \!\!\! {p} = 
 \left( 
 \begin{tabular}{cc}
 $(p_0 + i p_3) 1_{2\times 2}$  &  $-\epsilon_{kl}p_k \sigma_l$   \\ 
 $\epsilon_{kl}p_k \sigma_l$    &  $(p_0 - i p_3) 1_{2\times 2}$  \\ 
 \end{tabular}
 \right) \;,
\ee
where $\epsilon_{kl}$ is antisymmetric 
and $\epsilon_{12} = 1$. Thus, denoting 
\be
  \psi \equiv \left( 
  \begin{array}{l} \chi \\ \phi \end{array}
  \right) \;,
\ee
where $\chi, \phi$ are two-component spinors, 
the Lagrangian in~\eq\nr{Lpsi} becomes
\be
 \mathcal{L}_E^\psi  = 
 i \chi^\dagger \Bigl[ p_0 - g A_0 + D_3 \Bigr] \chi + 
 i \phi^\dagger \Bigl[ p_0 - g A_0 - D_3 \Bigr] \phi
 +\phi^\dagger \epsilon_{kl} D_k \sigma_l \chi
 -\chi^\dagger \epsilon_{kl} D_k \sigma_l \phi 
 \;.  \la{qcd_basis}
\ee
Leaving out, for brevity, the flavour structures, 
the bilinear operators of \eqs\nr{Sa}--\nr{Aa} become
\ba
 S 
 & = & 
 \chi^\dagger \phi + \phi^\dagger \chi
 \;, \la{op1} \\ 
 P 
 & = & 
    \chi^\dagger \sigma_3 \phi -
    \phi^\dagger \sigma_3 \chi
 \;, \\ 
 V_0 
 & = & 
 \chi^\dagger \chi + \phi^\dagger \phi
 \;, \\ 
 V_k 
 & = & 
 -\epsilon_{kl} (\chi^\dagger \sigma_l \phi - 
 \phi^\dagger \sigma_l \chi)
 \;, \\ 
 V_3 
 & = & 
 \chi^\dagger \chi - \phi^\dagger \phi
 \;, \\ 
 A_0 
 & = & 
   \phi^\dagger \sigma_3 \phi
  -\chi^\dagger \sigma_3 \chi 
 \;, \\ 
 A_k 
 & = & 
 -i (\chi^\dagger \sigma_k \phi + \phi^\dagger \sigma_k \chi)
 \;, \\ 
 A_3 
 & = & 
 -i (\chi^\dagger \sigma_3 \chi + \phi^\dagger \sigma_3 \phi)
 \;. \la{op2}
\ea
Note that the parity transformation, 
$\psi(\tau,\vec{x}) \to \gamma_0 \psi(\tau, -\vec{x})$, 
reads $\chi(\tau,\vec{x}) \to \phi(\tau,-\vec{x})$, 
$\phi(\tau,\vec{x}) \to \chi(\tau,-\vec{x})$ in this basis. 

In \se\ref{se:limits}, we saw that in the correlators we are 
interested in, the quarks are almost ``on-shell''. Let us therefore
consider \eq\nr{qcd_basis} for on-shell configurations,
in momentum space. 
We fix $\vec{p}_\perp$ and consider the functional 
dependence on $p_3$. In the free case,  the on-shell point
is at $p_0^2 + p^2 = 0$, i.e., 
$p_3 = \pm i [ p_0 + \vec{p}_\perp^2/(2 p_0) + ...]$. 
The positive and negative ``energy'' solutions could 
be called the particles and the anti-particles.

Because the quarks interact (at the tree-level) with bosonic 
Matsubara zero modes only, we expect that off-shellness is related
to the momentum scales of the latter: 
$|p_3\pm i p_0| \lsim gT$. 
Assuming this to be the case, the idea is to construct an effective theory
describing the soft scale dynamics. 
For instance,  
consider the pole around $p_3 \approx i p_0$. Then $\chi$ represents the
light mode, while $\phi$ is heavy. Being so far at the classical
level, we can solve for $\phi$ from the equations of motion, 
and substitute the solution back to the Lagrangian, resulting
in an action for $\chi$ alone  
(cf.,\ e.g.,\ the pedagogic presentation in~\cite{mn} for 
the 4d zero temperature case). 
The same treatment naturally
applies to $\phi$ around its pole. 
Expanding finally in powers of $1/p_0$, we obtain
a ``diagonalised'' on-shell
effective Lagrangian for two independent light modes, 
with a non-relativistic structure:
\ba
 \mathcal{L}_E^\psi & \approx & 
 i \chi^\dagger
 \Bigl[ 
 p_0 - g A_0 + D_3 - \frac{1}{2p_0}
 \Bigl(
 D_k^2 + \frac{g}{4i} [\sigma_k,\sigma_l] F_{kl} 
 \Bigr)
 \Bigr]
 \chi 
 \nn  
 & + & 
 i \phi^\dagger
 \Bigl[ 
 p_0 - g A_0 - D_3 - \frac{1}{2p_0}
 \Bigl(
 D_k^2 + \frac{g}{4i} [\sigma_k,\sigma_l] F_{kl} 
 \Bigr)
 \Bigr]
 \phi + \mathcal{O}\Bigl(\frac{1}{p_0^2}\Bigr) 
 \;.  
 \la{nrqcd3}
\ea 
The mesonic states are most usefully still represented
by the operators in~\eqs\nr{op1}--\nr{op2}.

For future reference, let us note that the free propagators
following from~\eq\nr{nrqcd3} are
\ba
 \langle \chi_u(p) \chi_v^*(q) \rangle & = & \delta_{uv}  
 (2\pi)^3 \delta^{(3)}(p-q) \frac{-i}{M + i p_3 
 + \vec{p}_\perp^2/(2 p_0)}
 \;, 
 \la{fullPq1}
 \\
 \langle \phi_u(p) \phi_v^*(q) \rangle & = & \delta_{uv}  
 (2\pi)^3 \delta^{(3)}(p-q) \frac{-i}{M - i p_3 
 + \vec{p}_\perp^2/(2 p_0)}
 \;, 
 \la{fullPq2}
\ea
where $M\equiv p_0$, and $u,v$ contain the spinor, flavour and colour 
indices. We may also remark that going into configuration space, 
the propagators become
\ba
 \langle \chi_u(x) \chi_v^*(y) \rangle \!\! &  = & \!\!
  - i \delta_{uv} 
      \theta(x_3 - y_3) 
  \frac{p_0}{2\pi |x_3 - y_3|} 
  \exp\Bigl({-M |x_3-y_3|
  -\frac{p_0(\vec{x}_\perp - \vec{y}_\perp)^2}{2 |x_3 - y_3|}}\Bigr)
  , \hspace*{0.5cm} 
  \la{statPx1} \\
  \langle \phi_u(x) \phi_v^*(y) \rangle \!\! & = & \!\! 
  - i \delta_{uv} 
      \theta(y_3 - x_3) 
  \frac{p_0}{2\pi |x_3 - y_3|} 
  \exp\Bigl({-M |x_3-y_3|
  -\frac{p_0(\vec{x}_\perp - \vec{y}_\perp)^2}{2 |x_3 - y_3|}}\Bigr)
  . \hspace*{0.5cm}
  \la{statPx2}
\ea
This implies forward time propagation for $\chi$ 
and backward for $\phi$. 
For $p_0\to \infty$, 
\be
  \frac{p_0}{2\pi |x_3 - y_3|} 
  \exp\Bigl({-\frac{p_0(\vec{x}_\perp - \vec{y}_\perp)^2}{2 |x_3 - y_3|}}\Bigr)
 \to \delta^{(2)}(\vec{x}_\perp - \vec{y}_\perp) \;, 
\ee
indicating that the quarks become static.

To conclude this Section, 
it may be noted that in~\eqs\nr{op1}--\nr{op2}
the charges ($V_0,A_0$) and the third components
of the currents ($V_3,A_3$)
are of the type $\sim \chi^\dagger \chi +  \phi^\dagger \phi$. This
implies that the corresponding correlators decay very rapidly:
according to \eqs\nr{statPx1}, \nr{statPx2}, 
they in fact vanish identically at non-zero distances, 
being of the form $\sim \theta(t)\theta(-t)$. 
For the third components of the currents this is
due to current conservation (the correlator is a constant, 
and vanishes in infinite volume, up to possible
contact terms), while for the
charges, this is related to the discussion in~\se\ref{se:limits}, 
where we observed  the prefactor $({q}^2 + 4 p_0^2)$ in the free 
correlators. In the on-shell limit, this causes the correlators 
to vanish; in reality, they are power-suppressed in $1/p_0 t$.
In the following, we will concentrate on the correlators which
decay more slowly, that is, which are represented by operators 
of the type $\sim \chi^\dagger \phi + \phi^\dagger \chi$.

\subsection{Power-counting and loop corrections}
\la{se:loop}

The construction of the effective
theory in~\eq\nr{nrqcd3} was so far on the tree-level. In order 
to be sure that the IR structure of the effective theory is really 
that of finite temperature QCD, and in order to determine the
radiative corrections we are ultimately interested in, we need 
to consider loop corrections. In particular, we should
find out the order to which one needs to go 
in the matching computation between the original
finite temperature QCD and NRQCD$_3$, as well as the resolution 
with which the dynamics inside NRQCD$_3$ needs to be treated, 
given that we are interested in the $\mathcal{O}(g^2)$ correction
to the spatial mesonic correlation lengths. 
For these purposes, it is useful to set up some 
parametric power-counting rules. 

Since the general upshot will be rather obvious, let us state
it immediately: in order to determine the
$\mathcal{O}(g^2)$ correction to the mesonic correlation lengths, 
the additive parameter $p_0$ in~\eq\nr{nrqcd3} becomes a matching 
coefficient (to be denoted by $M$, as in \eqs\nr{fullPq1}--\nr{statPx2}) 
which has to be determined
to 1-loop order. On the other hand, no other parameters need
to be matched at this level, and some interactions can even 
be dropped out: the final form of the theory is displayed 
in~\eq\nr{nrqcd_final} below. This matching takes care of the
``hard part'' in loop integrals of the type in~\fig\ref{fig:nrt}(a), while 
the dynamics within the effective theory accounts for diagrams of the 
type in~\fig\ref{fig:nrt}(b), where the gauge fields are always soft.  

To account for the full correlator defined in~\eq\nr{Cq}, one 
should also carry out operator matching to order $\mathcal{O}(g^2)$.
We will not need to consider it here, however, since we will not 
attempt to determine the overall magnitude of the correlator, only the 
coefficient of its exponential falloff.

Let us now turn to the power-counting rules. 
As mentioned, the parametric scales in the problem are 
$\sim \pi T,gT,g^2T$. Let us write $m\equiv \pi T, v\equiv g$, 
and denote by $\Delta p_3$ the off-shellness of the ``energy'', 
$\Delta p_3 = p_3 \pm i p_0$. 
Following the terminology of NRQCD,
one can define four different regions of phase space:
\ba
 \mbox{hard (h)}&: & \Delta p_3 \sim |\vec{p}_\perp| \sim m \;, \nn
 \mbox{soft (s)}&: & \Delta p_3 \sim |\vec{p}_\perp| \sim mv \;, \nn
 \mbox{potential (p)}&: & \Delta p_3 \sim mv^2\;, |\vec{p}_\perp|  
                           \sim mv \;, \nn
 \mbox{ultrasoft (us)}&: & \Delta p_3 \sim |\vec{p}_\perp| \sim mv^2 \;.   
 \la{regs}
\ea
By definition, the NRQCD$_3$ theory in~\eqs\nr{nrqcd3}, \nr{LA} 
is chosen to contain at most soft quarks and gluons as dynamical 
degrees of freedom. In fact, close to on-shell, the quarks of  
NRQCD$_3$ are potential, 
since $\Delta p_3 \sim |\vec{p}_\perp^2|/m \sim m v^2$. 

In order to estimate the importance of the 
various momentum regions in~\eq\nr{regs}, 
it is helpful to assign power-counting rules also to the corresponding
field components. This can be done by going to configuration space and
requiring that the action, a pure number, is parametrically of order unity. 
For soft and potential quarks, $\chi_\rmi{s},\chi_\rmi{p}$,
it follows from  
$S \sim \int {\rm d}t\, {\rm d}^2 \vec{x}_\perp 
\, \chi^\dagger i \partial_t \chi \sim 1$ 
that 
$\chi \sim 1/|\vec{x}_\perp| \sim mv$. 
For soft gluons $A_\rmi{s}$,  
$S \sim \int \! {\rm d}t\, {\rm d}^2 \vec{x}_\perp \, 
A_\rmi{s} \partial_t^2 A_\rmi{s} \sim 1$
implies that
$A_\rmi{s} \sim (t/\vec{x}_\perp^2)^{1/2} \sim m^{1/2}v^{1/2}$. 
For ultrasoft gluons $A_\rmi{us}$, correspondingly,  
$A_\rmi{us} \sim (t/\vec{x}_\perp^2)^{1/2} \sim m^{1/2}v$. 
In addition, when operating on on-shell quarks, the time 
derivative can be estimated as $\partial_t \sim 1/t \sim mv^2$. 

Given these rules and that 
$g_\rmi{E}^2 \sim m v^2$,
we note that  
$g_\rmi{E} A_\rmi{s} \sim mv^{3/2}$, $g_\rmi{E} A_\rmi{us} \sim mv^2$.
The important conclusion follows that soft and ultrasoft
gauge fields are parametrically
of {\em higher order} than the transverse
momentum $\nabla_\perp\sim mv$. 
In particular, given that we are interested in quarks which are off-shell 
by at most $\mathcal{O}(mv^2)\sim \mathcal{O}(g^2T)$, 
we should note that this magnitude is already
saturated by $\nabla_\perp^2/m$: 
no transverse covariant derivatives are needed in the quark action!
On the contrary, the gauge fields {\em are} significant in relation to 
$\partial_t$. 

To summarise, if we want to keep the Lagrangian up to and including
the order $\mathcal{O}(m^3 v^4)$, such that we know the on-shell quark
self-energy up to order $\mathcal{O}(mv^2)$, it is enough to 
replace~\eq\nr{nrqcd3} with the action
\be
 \mathcal{L}_E^\psi = 
 i \chi^\dagger \biggl(
 M - g_\rmi{E} A_0 + D_t 
 - \frac{\nabla_\perp^2}{2 p_0} 
 \biggr) \chi + 
 i \phi^\dagger \biggl(
 M - g_\rmi{E} A_0 - D_t 
 - \frac{\nabla_\perp^2}{2 p_0} 
 \biggr) \phi  
 \;,
 \la{nrqcd_eff}
 \la{nrqcd_final}
\ee 
where $D_t \equiv D_3 = \partial_3 - i g_\rmi{E} A_3$, 
and the gauge fields have now the same 3d normalisation as in~\eq\nr{LA}.
Furthermore, the only parameter requiring a matching computation is $M$. 

To conclude this Section, let us briefly view 
the dynamics of the remaining degrees of freedom.
Since the quarks are very heavy, of mass $p_0$, we may
compute the static potential in (2+1) dimensions. Its
structure is, 
\be
 V(r) \sim g_\rmi{E}^2 \ln r + g_\rmi{E}^4 r + \mathcal{O}(g_\rmi{E}^6 r^2) 
 \;.
 \la{pot0}
\ee 
If inserted into a Schr\"odinger equation, we get 
\be
 \frac{1}{p_0} \frac{\partial^2}{\partial r^2} \sim V(r)
 \sim g_\rmi{E}^2 \ln r \;,
\ee
and therefore $|\vec{p}_\perp|
\sim \partial/\partial r\sim 1/r \sim \sqrt{g_\rmi{E}^2 p_0}
\sim g T$. 
We see that
the expansion parameter related to the static potential
is $\mathcal{O}(g_\rmi{E}^2r) \sim \mathcal{O}(g)$. 
Consequently, for the correction of order 
$E \sim |\vec{p}_\perp|^2 /p_0 \sim g^2 T$, 
it is sufficient 
to compute the static potential to leading order. 
The leading order contribution itself may, 
however, already be infrared divergent, in the sense that the
scale appearing inside the logarithm in~\eq\nr{pot0}
remains to be determined. 

%
\section{Matching from QCD to NRQCD$_3$}
\la{se:match}

As explained above, the parameter we need to determine through
a matching computation is $M$, as defined by \eq\nr{nrqcd_eff}.
This computation was not carried out in~\cite{hl}, although many 
other matchings related to NRQCD$_3$ were. 

In order to avoid having to deal with any wave function 
normalisation factors, we will carry out the matching by 
computing the quark ``pole mass'', or finite temperature Euclidean
dispersion relation, both in the original finite temperature QCD, 
as well as with the theory of~\eq\nr{nrqcd_eff}. Both computations
produce a gauge-invariant result, and requiring the outcomes to be 
equivalent, leads to an expression for $M$ in terms of the parameters
of finite temperature QCD. 

In order to monitor explicitly the gauge parameter 
dependence and the IR sensitivity of the computations carried out, 
we take the gluon propagator to be of the form 
\be
 \langle\, A^a_\mu (p) A^b_\nu(q) \,\rangle
 = 
 \delta^{ab} (2\pi)^{4-2\epsilon} \delta^{(4-2\epsilon)}(p+q)
 \biggl[  
 \biggl(
 \delta_{\mu\nu} - \frac{p_\mu p_\nu}{p^2}
 \biggr) 
 \frac{1}{p^2 + \lambda^2} + 
 \frac{p_\mu p_\nu}{p^2}
 \frac{\xi}{p^2 + \xi \lambda^2}
 \biggr] 
 \;.
 \la{Amu}
\ee
That is, we introduce a gauge parameter $\xi$ as well as
an IR-cutoff $\lambda$. As we will see, the pole masses, computed
in both ways, are independent of $\xi$, and allow to take
the limit $\lambda \to 0$. The same is then also true for their
difference, determining the matching coefficient we are interested in. 

An important and somewhat non-trivial point is the way the computation
is to be carried out on the side of NRQCD$_3$. The ``problem'' is that
we want to use dimensional regularisation, yet obey the power-counting
conventions as discussed above. If one, however, takes the free
propagators as determined by~\eq\nr{nrqcd_eff}, shown 
in~\eqs\nr{fullPq1}, \nr{fullPq2}, then (irrespective of 
whether or not one keeps $M$ there explicitly or shifts
it away by modifying the integration contour for $p_3$) momenta 
of order $|\vec{p}_\perp|\sim p_0$ do contribute. This, however, 
we do not want, because our power-counting rules assumed the 
dynamical scales inside the effective theory to be at most soft, 
$|\vec{p}_\perp| \lsim m v$, while $p_0 \sim m$. 
The way to avoid the problem with dimensional regularisation
has been clarified in~\cite{am}: one should carry out the matching
in the soft region of the phase space, 
$p_3 \sim |\vec{p}_\perp|\sim mv$, where the 
transverse part $\vec{p}_\perp^2/p_0$ can be treated as a perturbation.
That is, we match the operators of NRQCD$_3$ order by order 
in $1/p_0$. 

The situation would be different if we were to treat NRQCD$_3$
also on the lattice. In order to match from dimensional regularisation
to lattice regularisation, say, 
one would then really have to compute using the same
forms of propagators in both schemes. This 
leads to additional logarithmic ultraviolet divergences in the parameter $M$, 
needed in order to cancel the corresponding divergences appearing
in the dynamics within NRQCD$_3$. Carrying out the computation 
this way is not much more complicated than what we do here, but 
since the results are not needed for our present purposes, 
we do not display them.

On the side of QCD, the inverse propagator evaluated at 
the position of the tree-level pole, $p^2 = 0$, reads 
\be
 \Sigma(p) = i \slash\!\!\!{p} - i g^2 C_F \left.
 \Tint{q} \frac{\gamma_\mu (\slash\!\!\!{p} 
 - \slash\!\!\!{q})\gamma_\mu}
 {(p-q)^2_f (q^2 + \lambda^2)_b} \right|_{p^2=0} \;,
\ee
where $(...)_f$ denotes fermionic Matsubara modes, $(...)_b$
bosonic, and $C_F = (N_c^2 - 1)/ 2 N_c$. 
The result comes from the transverse part of the gluon propagator; 
the longitudinal gauge parameter dependent part
does not contribute at the pole. 
The task is now to find the zeros of $\Sigma(p)$, 
to relative order $\mathcal{O}(g^2)$.

In order to streamline the task, we may set $\vec{p}_\perp = 0$, 
and multiply $\Sigma(p)$ from the left with $\gamma_0$. The result
obtains a simple form if we choose the basis of $\gamma$-matrices
introduced in~\se\ref{se:tree}. 
Indeed, off-diagonal terms vanish after integration
over $q$, and the result becomes block-diagonal. Moreover, the upper
two components are degenerate, as well as the lower two components, 
and the latter follow from the former 
simply by $(p_3 \to -p_3, q_3\to -q_3)$. Therefore
it is enough to consider
\ba
 \Bigl[ \gamma_0 \Sigma(p) \Bigr]_{11} = 
 i \Bigl\{ 
 p_0 + i p_3 - g^2 C_F \Tint{q} 
 \frac{1}{(p-q)^2_f (q^2 + \lambda^2)_b} \Bigl[ 
 && \hspace*{-0.7cm}  + p_0 - i p_3 - q_0 + i q_3  \quad \quad \quad (A_0) 
 \la{Sp} \\
 && \hspace*{-0.7cm}  
 - p_0 + i p_3 + q_0 - i q_3  \quad \quad \quad  (A_3) 
 \nn 
 && \hspace*{-0.7cm} 
 \raise-1ex
 \hbox{$-2 p_0 - 2 i p_3 + 2 q_0 + 2 i q_3  \quad (A_k)
\Bigr]\Bigr\} \;,$} \nonumber
\ea
where the gauge field
components leading to the various contributions have also been indicated, 
and we have set the spatial dimensionality 
to $d=3$, anticipating the fact that
the result is UV-finite. 
The $\mathcal{O}(g^2)$ term is to be evaluated 
at the tree-level pole position $p_0 = -i p_3$. 

It can be observed from~\eq\nr{Sp} that 
both $A_0$ and $A_3$
give possibly logarithmically IR-divergent integrals (if $\lambda\to 0$),
from the terms with $p_0 - i p_3$ in the numerator, but the 
sum is finite since they come with opposite signs. 
For $A_k$, on the other hand, the contribution involving $p_0, ip_3$ comes
in the combination $p_0 + i p_3$, and vanishes altogether 
at the pole. The correction as a whole is therefore IR finite,
and can be evaluated with $\lambda \to 0$, 
whereby the integral becomes
\be 
 \left.
 \Tint{q} \frac{2(q_0 + i q_3)}{(p-q)^2_f (q^2)_b}
 \right|_{p_0 = -i p_3}
 = 
 \left.
 \frac{1}{p_0}
 \Tint{q} \frac{2 p \cdot q}{(p-q)^2_f (q^2)_b}
 \right|_{p^2 = 0}
 = \frac{1}{p_0} \Tint{q}
 \biggl[ 
 -\frac{1}{q_b^2} + \frac{1}{q_f^2}
 \biggr]  = -\frac{T^2}{8 p_0} \;.
\ee
Consequently, 
\be
  \Bigl[ \gamma_0 \Sigma(p) \Bigr]_{11} \approx
  i \Bigl\{ p_0 + i p_3 + g^2 C_F \frac{T^2}{8 p_0}
  \Bigr\} \;,
  \la{matchQCD} 
\ee 
and the dispersion relation, or pole position, becomes
\be
 p_3 \approx i \Bigl[
   p_0 + g^2 C_F \frac{T^2}{8 p_0} 
 \Bigr] \;.
 \la{disp}
\ee

On the side of NRQCD$_3$ we proceed  
as explained above, following~\cite{am}. 
The gauge field propagator
is treated as in~\eq\nr{Amu} at the current stage 
(during matching, any possible IR cutoffs should be the same 
in the two theories).
The inverse propagator now becomes
\ba 
 \hat \Sigma(p) & = &  
 i
 \biggl\{
  M +  i p_3 - g_\rmi{E}^2 C_F 
 \int \frac{{\rm d}^{3-2\epsilon} q}{(2\pi)^{3-2\epsilon}} 
 \frac{1}{M + i p_3 - i q_3} 
 \biggl[ + \frac{1}{q^2 + \lambda^2} \quad\quad (A_0) 
  \la{matchNR} \\
 & & 
 \hphantom{i
 \biggl\{
  M + i p_3 
  - g_3^2 C_F 
 \int \frac{{\rm d}^{3-2\epsilon} q}{(2\pi)^{3-2\epsilon}} 
 \frac{1}{M +i p_3 - i q_3}} 
 - \frac{1}{q^2 + \lambda^2} \quad\quad (A_3)
 \biggr] 
 \biggr\}
 \;.
 \nonumber
\ea
The potential IR divergences from the two gauge field components are
the same as in~\eq\nr{Sp},
but they cancel, as they did there, such that
the 1-loop correction is finite, and in fact vanishes. 
Thus, the dispersion relation is simply
\be
 p_3 \approx i M \;. 
 \la{dispNR}
\ee
Matching now \eqs\nr{disp}
and \nr{dispNR}, we obtain 
\be
  M = p_0 + g^2 T \frac{C_F}{8\pi} + \mathcal{O}(g^4 T)
  \;. 
  \la{finalM}
\ee

%
\section{Perturbative solution within NRQCD$_3$}
\la{se:potential}

After the construction of the effective theory, which accounts
for hard gluons in~\fig\ref{fig:nrt}(a), we still 
have to account for the soft part of~\fig\ref{fig:nrt}(a) as well as
for all of~\fig\ref{fig:nrt}(b), where the gluons can only be soft (i.e., 
Matsubara zero modes). This is to be done by computing the analogues 
of the correlators in~\eq\nr{Ct}, using the effective theory in 
\eqs\nr{LA}, \nr{nrqcd_eff}, and the operators 
in~\eqs\nr{op1}--\nr{op2}. Note that because of the appearance
of Debye screening (represented by the parameter $m_\rmi{E}^2$
in~\eq\nr{LA}), the gauge field components $A_0, A_3$  
behave now differently: the propagators are
\ba
 \langle A^a_0 (p) A^b_0(q) \rangle
 & = & 
 \delta^{ab} (2\pi)^{3-2\epsilon} 
 \delta^{(3-2\epsilon)}(p+q)\frac{1}{p^2 + m_\rmi{E}^2} 
 \;, \\
 \langle A^a_i (p) A^b_j(q) \rangle
 & = & 
 \delta^{ab} (2\pi)^{3-2\epsilon} \delta^{(3-2\epsilon)}(p+q)
 \biggl[  
 \biggl(
 \delta_{ij} - \frac{p_i p_j}{p^2}
 \biggr) 
 \frac{1}{p^2 + \lambda^2} + 
 \frac{p_i p_j}{p^2}
 \frac{\xi}{p^2 + \xi \lambda^2}
 \biggr] 
 \;. \hspace*{0.5cm}
\ea
We have again introduced a fictitious infrared regulator $\lambda$, 
as well as a general gauge parameter $\xi$, just in order to check that 
the results are independent of them. 

The correlators of~\eq\nr{Ct} take the form
\be
 C_{t}[O^a,O^b] \sim 
 \int \! {\rm d}^2 \vec{x}_\perp \, 
 \Bigl \langle
 O^a(\vec{x}_\perp,t) O^b(\vec{0}_\perp,0)
 \Bigr \rangle
 \;,
 \la{C_red} 
\ee
where we have suppressed the dependence on the $\tau$-coordinate
which disappeared as we moved to Matsubara frequency space.  
As already mentioned, we are ignoring corrections of relative order
$\mathcal{O}(g^2)$ in the overall normalisation of~\eq\nr{C_red}, but
not in the coefficient of the exponential falloff.  

Even though the operators in~\eqs\nr{op1}--\nr{op2} 
contain two terms, $\sim \chi^\dagger \phi+ \phi^\dagger \chi$, 
only one product contributes
in~\eq\nr{C_red}: this is the one where both quarks are forward 
propagating (cf.\ \eqs\nr{statPx1},
\nr{statPx2}), $\sim \langle 
\phi^\dagger(t) \chi(t) \, \chi^\dagger(0) \phi(0) \rangle$. 
In other words, we need to determine the Green's function for the 
composite object 
$\int \! {\rm d}^2 \vec{x}_\perp \, 
\phi_u^*(\vec{x}_\perp,t) \Gamma_{uv} \chi_v(\vec{x}_\perp,t)$, 
where $\Gamma_{uv}$ is a $2\times 2$ matrix in spinor
space, an $N_f\times N_f$ matrix in flavour space,
and an $N_c \times N_c$ matrix in colour space. 
Since the spinor and flavour indices play no role in the 
following, they will be suppressed.

As we have argued in~\se\ref{se:loop}, 
it should be rather obvious how to proceed now:
simply integrate out the gauge fields, to produce 
a static potential for the $\phi^*\chi$ pair, 
and solve the corresponding Schr\"odinger equation. The reason this
works is that, as we will show {\it a posteriori}, the non-perturbative 
ultrasoft gauge fields do not contribute at the present order.
The problem could also be 
formalised for instance by
going through the PNRQCD effective theory~\cite{pnrqcd,mb}, from
which the soft gluonic scales have been systematically integrated out. 
For our modest purposes here, involving only the Coulomb-type potential,
going through the full
formalism of PNRQCD seems somewhat too elaborate, however. What
we rather wish to do is to recall on a concrete but somewhat
heuristic level how the Schr\"odinger equation can be seen to emerge, 
just in order to get all the signs right in our final expression. 

The Green's function we need to consider is~\eq\nr{C_red}. 
We would like to find the equation of motion obeyed by this 
Green's function, which we expect, at least
at large $t$, to be of the form $(\partial_t - H) G(t) = C \delta(t)$,
where $H$ is a differential operator and 
$C$ is a constant. Then the exponential decay is 
determined by the lowest eigenvalue of $H$.

In order to find $H$, 
we now define a point-splitting (which is essentially gauge-invariant 
because no spatial components of gauge fields
appear in~\eq\nr{nrqcd_eff}), by introducing
a vector $\vec{r}$ and, with the intention of taking 
$\vec{r}\to 0$ afterwards, consider the modified correlator
\ba
 C(\vec{r},t) & \equiv & 
 \int \! {\rm d}^{2-2\epsilon} \vec{R} \, 
 \Bigl\langle
 \phi^*(\vec{R}+\vec{r}/2,t)
 \chi(\vec{R}-\vec{r}/2,t) 
 \;
 \chi^* (\vec{0},0)
 \phi(\vec{0},0) 
 \Bigr\rangle
 \;.
\ea
We denote the tree and 1-loop contributions 
by $C^{(0)}(\vec{r},t), C^{(1)}(\vec{r},t)$. It is then easy to see that
\be
 \Bigl[
 \partial_t + 2 M - \frac{1}{p_0} \nabla_\vec{r}^2 
 \Bigr] C^{(0)}(\vec{r},t) \propto \delta(t)\delta^{(2-2\epsilon)}(\vec{r})
 \;.
 \la{CA}
\ee
The result for the 1-loop contribution 
(a sum of the first two graphs in~\fig\ref{fig:nrt})
can, on the other hand, be written in the form
\be
 \Bigl[
 \partial_t + 2 M - \frac{1}{p_0} \nabla_\vec{r}^2 
 \Bigr] C^{(1)}(\vec{r},t) = 
 - g_\rmi{E}^2 C_F p_0^{-2\epsilon} \, 
 \mathcal{K} \biggl( \frac{1}{t p_0}, \frac{\nabla_\vec{r}}{p_0},
 \frac{m_\rmi{E}^2}{p_0^2}, \frac{\lambda^2}{p_0^2}, \vec{r} p_0
 \biggr)
  C^{(0)}(\vec{r},t)
 \;, 
 \la{CB}
\ee
where the kernel $\mathcal{K}$ is dimensionless.
Since the kernel is multiplied
by $g_\rmi{E}^2$, we may assume that it can be expanded with respect
to its first two arguments, given that non-trivial terms in this
expansion should be subleading in $g_\rmi{E}^2/p_0$ with respect to the terms 
$\partial_t , -\nabla_\vec{r}^2/p_0$ already appearing in~\eq\nr{CA}. 
The zeroth order term
turns out to involve a momentum space delta-function $\delta(q_3)$, which
restricts the remaining loop integral to be
($2-2\epsilon$)-dimensional, and produces
just the 1-loop static potential,
\ba
 V(\vec{r}) & \equiv &
 g_\rmi{E}^2 C_F p_0^{-2\epsilon}\, \mathcal{K}\biggl( 0,0,
 \frac{m_\rmi{E}^2}{p_0^2}, \frac{\lambda^2}{p_0^2}, \vec{r} p_0
 \biggr)
 \nn 
 & = &  
 g_\rmi{E}^2 C_F \int \frac{{\rm d}^{2-2\epsilon} q}
 {(2\pi)^{2-2\epsilon}} \biggl\{
 \frac{1}{q^2 + \lambda^2} \Bigl[ 1 - e^{i \vec{q}\cdot\vec{r}}\Bigr]
 - \frac{1}{q^2 + m_\rmi{E}^2} \Bigl[ 1 + e^{i \vec{q}\cdot\vec{r}}\Bigr]
 \biggr\} \;.
 \la{Vr}
\ea 
Writing now $C(\vec{r},t) = C^{(0)}(\vec{r},t) + C^{(1)}(\vec{r},t)$, 
we can combine \eqs\nr{CA}, \nr{CB}, \nr{Vr} to
\be
 \Bigl[
 \partial_t + 2 M - \frac{1}{p_0} \nabla_\vec{r}^2 + V(\vec{r})
 \Bigr] C (\vec{r},t) \propto \delta(t)\delta^{(2-2\epsilon)}(\vec{r})
 \;, 
 \la{H}
\ee
where the point-splitting can be removed, 
$\vec{r}\to 0$, after the solution, and subleading terms have been dropped. 
Therefore, the exponential falloff is determined by 
the eigenvalues of the Hamiltonian read from \eq\nr{H}.

It is worthwhile to note, in passing, the signs appearing
in the two terms in~\eq\nr{Vr} (the first from $A_3$, the second
from $A_0$). The reason for the differences are the ways in which
$A_3,A_0$ appear in~\eq\nr{nrqcd_final}: $\pm i g_\rmi{E} A_3$ versus
$-g_\rmi{E} A_0$. The fact that the $\vec{r}$-independent terms from $A_3,A_0$
come with opposite signs, was already met in \eqs\nr{Sp}, \nr{matchNR}. 

Employing the basic integrals
\ba
 \int \frac{{\rm d}^{2-2\epsilon}q}
 {(2\pi)^{2-2\epsilon}} \frac{1}{q^2 + \lambda^2} 
 & = & \frac{\mu^{-2\epsilon}}{4\pi}
 \biggl( 
 \frac{1}{\epsilon} + 2 \ln\frac{\bmu}{\lambda} + \mathcal{O}(\epsilon)
 \biggr) \;, \\
 \int \frac{{\rm d}^{2}q}
 {(2\pi)^{2}} \frac{e^{i \vec{q}\cdot\vec{r}}}{q^2 + \lambda^2} 
 & = & 
 \frac{1}{2\pi} K_0(\lambda r)\;, 
\ea
where $\bmu$ is the $\msbar$ scale parameter, 
$\mu = \bmu (e^\gamma/4\pi)^{1/2}$, 
and $K_0$ is a modified Bessel function, 
as well as the asymptotic behaviour of $K_0$, 
\be
 K_0(x) = -\ln \frac{x}{2} - \gamma_E + \mathcal{O}(x) \;, 
\ee
it is observed that \eq\nr{Vr} is both ultraviolet finite 
($\lim_{\epsilon\to 0}$ exists) and infrared finite 
($\lim_{\lambda\to 0}$ exists). Its final form is
\be
 V(\vec{r}) = g_\rmi{E}^2 \frac{C_F}{2\pi}
 \Bigl[
 \ln\frac{m_\rmi{E} r}{2} + \gamma_E - K_0 (m_\rmi{E} r) 
 \Bigr] \;.
 \la{Vr2}
\ee
The potential determines the coefficient of the exponential fall-off, 
$\xi^{-1} \equiv m_\rmi{full}$, through
\be
 \Bigl[2 M  
 -\frac{\nabla_\vec{r}^2}{p_0} + V(\vec{r})
 \Bigr] \Psi_0 = m_\rmi{full} \Psi_0 \;,
 \la{se} 
\ee
where $\Psi_0$ is the ground state wave function. Therefore, 
$m_\rmi{full}$ is indeed insensitive to the gluonic ultrasoft 
scale at this order: there are no ambiguities inside the 
logarithm in~\eq\nr{pot0}.

In order to determine $m_\rmi{full}$, 
we have to solve \eq\nr{se}. We are not 
aware of an analytic solution for this bound state problem. In order 
to find the solution numerically, we rescale 
\be 
 r \equiv \hat \frac{r}{m_\rmi{E}},
 \quad 
 m_\rmi{full} - 2 M \equiv g_\rmi{E}^2 \frac{C_F}{2\pi} \hat E_0\;,
 \la{Ehat}
\ee
whereby~\eq\nr{se} becomes
\be
 \biggl[
 -\biggl(
 \frac{{\rm d}^2}{{\rm d}\hat r^2} + 
 \frac{1}{\hat r} \frac{{\rm d}}{{\rm d}\hat r} 
 \biggr) 
 + \rho
 \biggl( 
 \ln \frac{\hat r}{2} + \gamma_E - K_0(\hat r) - \hat E_0
 \biggr)
 \biggr] \Psi_0 = 0\;,
 \la{normS}
\ee
where
\be
 \rho = \frac{p_0 g_\rmi{E}^2 C_F}{2\pi m_\rmi{E}^2} = 
 \frac{3(N_c^2 - 1)}{2 N_c (2 N_c + N_f)} = 
 \left\{ 
 \begin{tabular}{ll}
 $2/3$   & , $N_f = 0$ \\
 $1/2$   & , $N_f = 2$ \\
 $4/9$   & , $N_f = 3$ 
  \end{tabular}
 \right. 
 \quad .
\ee

A numerical solution of~\eq\nr{normS}
is facilitated by determining the analytic 
behaviour of $\Psi_0(\hat r)$ around the origin. Assuming $\Psi_0(0)$
to be finite, one finds
\be
 \Psi_0(\hat r) \approx
 \Psi_0(0) \Bigl[ 
 1 + \fr12 \rho \, {\hat r}^2 \Bigl( 
 \ln\frac{\hat r}{2} + \gamma_E - 1 - \fr12 \hat E_0
 \Bigr)
 \Bigr] \;.
\ee
Integrating towards large $\hat r$ and requiring square integrability,
gives then 
\ba
 \hat E_0 & = &  0.16368014, \quad \rho = 2/3 \;, \quad (N_f = 0) 
  \la{Enum0}
  \\
  & = &  0.38237416, \quad \rho = 1/2 \;, \quad (N_f = 2) \\
  & = &  0.46939139, \quad \rho = 4/9 \;. \quad\, (N_f = 3)
  \la{Enum}
\ea
Therefore, for $N_c = 3$, $N_f = 0$ (corresponding to quenched QCD),   
\eqs\nr{finalM}, \nr{Ehat} and \nr{Enum0} together imply that
\be
 m_\rmi{full} = 2 \pi T + g_\rmi{E}^2 \frac{C_F}{2\pi} 
 \Bigl( \fr12 + \hat E_0 \Bigr)
 \approx  2 \pi T + 0.14083730\, g^2 T
 \;. \la{mfull}
\ee
It is worthwhile to note that this expression contains no logarithms. Here 
we differ from~\cite{hz}, for example, where the contributions from $A_0$ 
were left out. As can be seen in \eqs\nr{Vr}, \nr{Vr2}, these contributions
are essential.

At $T \sim 2 T_c$ for $N_f = 0$, 
it is estimated~\cite{adjoint} that 
the effective gauge coupling is $g_\rmi{E}^2/T \approx 2.7$. 
Thus the second term in~\eq\nr{mfull} represents a 
$\sim 5$\% correction to the leading order result, 
with a positive sign. Since the spinor structure of the operators
in~\eqs\nr{op1}--\nr{op2} played no significant role in the analysis
above (it only leads to an overall factor, $\tr 1$ or $\tr \sigma_3^2$), 
the same result holds for all the flavour non-singlet operators
of the type $\phi^\dagger \chi$, that is, $S^n,P^n,V_k^n,A_k^n$. 
As we have mentioned, the correlators
related to $V_0^n,A_0^n$ are suppressed, 
while those related to $V_3^n,A_3^n$ 
are constants because of current conservation, and 
vanish in infinite volume, such that no meaningful long-distance
screening masses can be associated with them.

%
\section{Comparison with lattice}
\la{se:result}

Let us now consider lattice measurements of the correlation
lengths discussed,  at $T > T_c$. This
is a classic problem in lattice QCD, initiated already 
a long time ago~\cite{dtk}. 
Most measurements concentrate on the non-singlets $P^n$, $V_\mu^n$
(or ``$\pi$, $\rho$''), because the singlets require an additional 
``disconnected''  contraction (corresponding to the quark lines 
in~\fig\ref{fig:nrt}(c)), whose inclusion is non-trivial. 
For recent original work on the topic, 
we refer to~\cite{recent}--\cite{ggm}, 
and for a recent review, to~\cite{lprev}.

As always with lattice simulations, practical measurements are 
faced with finite volume and discretisation artifacts.  
The measurements we are concerned with face also further problems, 
since the operators are fermionic: we would like to approach the chiral 
limit and simultaneously respect the flavour symmetries, which turn
out to be hard requirements to satisfy. 
Simulations could be carried out with, say, staggered, Wilson, domain 
wall, or overlap quarks, but in all cases but the last, which is 
very expensive in practice, it is oftentimes difficult to estimate
how serious the systematic errors are.
Finally, most lattice simulations are ``quenched'', ignoring the 
fermion determinant, which we however do not need to consider a
restriction, since we can easily set $N_f = 0$ in our predictions. 

To summarise the current status, from quenched measurements with Wilson
quarks at $N_\tau = 8$~\cite{recent}: already at $T \sim 1.5 T_c$, 
the $\pi$ is just slightly below $\rho$.
Both have masses consistent with or very slightly below the free 
theory prediction, that is $2\pi T$ in the continuum.
Very close to $T_c$ the situation has to be different, of course, 
and the pion mass should even vanish at $T_c$ while approaching the 
chiral limit with dynamical quarks, for $N_f = 2$. 

This pattern is quite different from what it used to be: in 
the early days, staggered quarks gave an anomalously low screening
mass for the pion even at $T > T_c$, both in the quenched as well 
as in the unquenched case. It has been found out, however, that 
discretisation effects are substantial for staggered quarks: 
in the quenched theory, 
the light masses increase in going from $N_\tau = 4$ to $N_\tau =12$, 
at $T = (1.5...2.0)T_c$, by up to $\sim$ 40\% (when masses are
expressed in units of $T$)~\cite{ggl,gg}. On the other hand, 
the effects of quenching seem to be moderate, although only tested 
far from the continuum limit (for $N_\tau =4$, staggered quarks)~\cite{ggm}.

To summarise, within the current resolution the mesonic screening
masses agree with the leading order prediction, within statistical
and certainly within systematic errors. It will require a 
significantly higher resolution to show unambiguously whether the
5\% correction we have found is really reproduced by the lattice 
data in the chiral continuum limit. In any case, 
the classic problem, that the 
$\pi$ and $\rho$ seemed not to be degenerate at $T > T_c$ in 
contrast to the perturbative prediction, has disappeared while 
approaching the continuum limit.

%
\section{Flavour singlets}
\la{se:singlets}

Flavour singlet quark bilinears behave in a different way than 
flavour non-singlets, because their correlation functions contain 
a ``disconnected'' contribution, of the type in \fig\ref{fig:nrt}(c).
This means that the operators couple, in general, to bosonic glueball states. 
They are also more difficult to measure on the lattice, just because
of the disconnected contraction. Here we discuss some basic features
concerning the flavour singlet correlators. 

The relations of the flavour singlets to gluonic operators can be 
determined by coupling the flavour singlets to sources in the Lagrangian, 
and then performing the Grassmann integral over the quarks, in a given
gauge field background. At high temperatures, this procedure is 
infrared safe, and the results can be expanded in $m^2/(\pi T)^2$.
To leading order in the expansion, a 1-loop computation of 0-gluon, 
1-gluon, 2-gluon
and 3-gluon vertices leads to 
\ba
 \frac{1}{N_f} \bar\psi \psi 
 & \leftrightarrow & 
 N_c \frac{mT^2}{6} - \frac{m}{2\pi^2} g^2 \tr A_0^2
 \;, \la{Ss} \\
 \frac{1}{N_f} \bar\psi \gamma_5 \psi 
 & \leftrightarrow & 
 ~ \frac{7\zeta(3) m}{8\pi^2 T^2} \frac{g^2}{32\pi^2} 
 \epsilon_{\alpha\beta\gamma\delta} F^a_{\alpha\beta} F^a_{\gamma\delta}
 \;, \la{g5} \la{Ps} \\
 \frac{1}{N_f} \bar\psi \gamma_0 \psi 
 & \leftrightarrow & 
 - \frac{i}{3\pi^2} g^3 \tr A_0^3
 \;, \la{nB} \\
 \frac{1}{N_f} \bar\psi \gamma_i \psi 
 & \leftrightarrow & 
 0 \;. \la{sVi} 
\ea
These relations imply, 
for instance, that the correlation length of $S^s=\bar\psi \psi$
can be determined from $\tr A_0^2$, and the 
correlation length of the baryon 
density~$V_0^s = \bar\psi\gamma_0 \psi$ from $\tr A_0^3$~\cite{mu,mu_t}. 
Furthermore, from
\ba
 \frac{g^2}{32\pi^2} 
 \epsilon_{\alpha\beta\gamma\delta} F^a_{\alpha\beta} F^a_{\gamma\delta}
 & = & \partial_\mu K_\mu \;, \\
 K_\mu & = & 
 \frac{g^2}{8\pi^2} \epsilon_{\mu\nu\lambda\rho}
 \biggl[
 A^a_\nu \partial_\lambda A^a_\rho + 
 \frac{g}{3} f^{abc} A^a_\nu A^b_\lambda A^c_\rho 
 \biggr] \;, 
\ea
it follows by integrating over $\int\! {\rm d}\tau\, {\rm d}^2\vec{x}_\perp$ 
that the correlation length of $P^s = \bar\psi \gamma_5 \psi$ is determined
by that of $K_3 \propto g^2 \epsilon_{ij3} A_0^a F_{ij}^a$, 
where the form indicated applies in the static limit. 
All of these gluonic correlation lengths
have been measured, for various $N_f$, in ref.~\cite{mu}.
It may also be noted from~\eqs\nr{Ss}, \nr{Ps}
that some of the couplings vanish in the chiral limit.

For the axial charge $A_0^s = \bar\psi\gamma_0\gamma_5\psi$, 
the relation to gluonic objects is 
very non-trivial because of the axial anomaly. We recall here
only that if one couples the axial charge to a constant chemical
potential, and then integrates over spacetime, one obtains~\cite{rw}
\ba
 \frac{1}{N_f} \bar\psi \gamma_0 \gamma_5 \psi 
 & \leftrightarrow & 
 K_0  \;. \la{Ax} 
\ea 
As is well-known 
this relation does not unambiguously apply for local operators, 
however (the right-hand-side, the Chern-Simons number density,
is not even gauge-invariant). Nevertheless it shows explicitly 
the quantum numbers of the gluonic objects that $A_0^s$ may couple to.

Let us finally discuss the spatial components of the 
vector and axial currents, 
and also touch the issue of U$_\rmi{A}$(1) axial 
symmetry non-restoration (for a review see, e.g., \cite{mm}).
To start with, we may recall that 
$\bar \psi \gamma_\mu \gamma_5 \psi$ and $\bar \psi \gamma_5 \psi$
are related through the anomaly equation 
\be
 \partial_\mu  [ \bar \psi \gamma_\mu \gamma_5 \psi ]
 = 2 m \, \bar \psi \gamma_5 \psi 
 + N_f \frac{g^2}{32\pi^2} 
 \epsilon_{\alpha\beta\gamma\delta} F^a_{\alpha\beta} F^a_{\gamma\delta}
 \;. 
 \la{anomaly}
\ee
Indeed, this relation remains true even at finite temperatures~\cite{im}, 
in spite of the fact that the ways in which 
$\partial_\mu  [ \bar \psi \gamma_\mu \gamma_5 \psi ]$ and
$\bar \psi \gamma_5 \psi $ separately couple to 
gluonic operators, change~\cite{cl}. 
Integrating now (the chiral $m\to 0$ limit of) \eq\nr{anomaly} 
over $\int \! {\rm d}\tau {\rm d}^2\vec{x}_\perp\,$ and noticing 
that the partial derivatives $\partial_3$ appearing on both 
sides do not modify the qualitative structure
and can thus be dropped out, 
it is observed that the 
correlators in the $x_3$-direction 
of $A^s_3 = \bar \psi \gamma_3 \gamma_5 \psi$ are determined 
by the operator $K_3 \propto g^2 \epsilon_{ij3} A_0^a F_{ij}^a$. 
Thus the correlator of $A_3^s$
falls off exponentially, with a correlation 
length determined~\cite{ay,lp} to be 
$\xi^{-1}[A_3^s] \equiv m[A_3^s] = m[K_3] \approx m_\rmi{E} + 
g_\rmi{E}^2 N_c (\ln(m_\rmi{E}/g_\rmi{E}^2)+7.0)/ 4\pi +\mathcal{O}(g^3 T)$. 
On the contrary, the vector current
is still conserved, and thus the correlator related to
$V^s_3  = \bar \psi\gamma_3 \psi$ is a constant, 
and vanishes in infinite volume.  
Thus, the long distance screening masses 
related to the U$_\rmi{A}$(1) partners $A_3^s,V_3^s$ 
are different even at arbitrarily high
temperatures and, in this sense, the axial U$_\rmi{A}$(1) symmetry 
is strictly speaking never restored. A similar conclusion of course
also holds, for instance, for $S^s,P^s$, as discussed above.

%
\section{Conclusions}
\la{se:concl}

The purpose of this paper has been to discuss analytic predictions
for the screening masses related to various quark bilinears at high
temperatures. We have addressed both flavour singlets, for which 
the screening masses are determined by those measured with bosonic 
glueball operators in the dimensionally reduced theory, and 
flavour non-singlets, for which we have determined the 
next-to-leading-order perturbative correction, \eq\nr{mfull}.

The next-to-leading-order correction we find
is small, but positive in sign. This is in contrast
to what lattice measurements traditionally indicated, and leads
to a non-monotonic behaviour for the mesonic screening masses,
whereby they first ``overshoot'' the perturbative result $2\pi T$
as the temperature is increased above $T_c$, and then approach
it from above for $T \gg T_c$. It will be
interesting to see whether this kind of a behaviour can be 
reproduced, as the lattice 
studies approach the infinite volume, continuum, and chiral limits. 
We should stress that our predictions hold formally also for the quenched
theory, probably the most immediate case on which 
progress could be expected. 

Apart from 4d lattice simulations, 
another interesting direction
for further study might be to
implement the effective theory we have derived, NRQCD$_3$, 
in lattice regularisation, and measure the correlation lengths
that way, following~\cite{tl} (assuming that the non-trivial
issues of renormalization and numerical instabilities in NRQCD$_3$ 
can be dealt with). This would allow to account for
some higher order corrections as well as non-perturbative effects, 
showing in particular whether those could change the sign or the magnitude
of the correction we have discussed, at realistic temperatures
not far above the deconfinement phase transition, where the coupling
constant is not asymptotically small. 

The ultimate theoretical goal of these studies is to understand
the dynamics of QCD at temperatures above a few hundred MeV, relevant 
for cosmology as well as for heavy ion collision experiments. 
One may ask, in particular, whether the strict
coupling constant expansion
might work for some observables after all, when carried out to a sufficient
depth (possibly including non-perturbative coefficients), 
in spite of the fact that
the first few corrections are huge~\cite{az,zk,bn,adjoint}. This would
allow to avoid the use of more complicated and 
partly phenomenological tools (for which many alternatives
are available, however~\cite{rvs}).  
Apart from gluonic observables~\cite{a0cond}--\cite{chris}, 
there are promising signs for this
also from related fermionic observables, 
quark number susceptibilities~\cite{av,av2}, but our case is more
infrared sensitive and in this sense a stronger test.

%
\section*{Acknowledgements}


We thank K.~Kajantie, E.~Laermann, and O.~Philipsen for 
discussions. 
M.V. was supported in part by the Finnish Cultural Foundation.


\appendix
\renewcommand{\thesection}{Appendix~\Alph{section}}
\renewcommand{\thesubsection}{\Alph{section}.\arabic{subsection}}
\renewcommand{\theequation}{\Alph{section}.\arabic{equation}}



\end{document}